\documentclass[twocolumn,preprintnumbers,amsmath,amssymb,floatfix,superscriptaddress,nofootinbib]{revtex4}
\usepackage{amsmath,hyperref,amssymb}
\usepackage{graphicx}
\usepackage{dcolumn}
\usepackage{bm}
\usepackage{verbatim}
\usepackage{tabularx}
\usepackage{slashed}
\usepackage{cancel}
\usepackage{hyperref}

\def\be{\begin{equation}}
\def\ee{\end{equation}}
\def\bea{\begin{eqnarray}}
\def\eea{\end{eqnarray}}
\def\ba#1\ea{\begin{align}#1\end{align}}
\def\bg#1\eg{\begin{gather}#1\end{gather}}
\def\bm#1\em{\begin{multline}#1\end{multline}}
\def\bmd#1\emd{\begin{multlined}#1\end{multlined}}

\newcommand{\mc}{\mathcal}
\renewcommand{\t}{\tilde}

\begin{document}

\title{Nonsupersymmetric dualities from mirror symmetry}

\author{Shamit Kachru}
\affiliation{\small \it Stanford Institute for Theoretical Physics, Stanford University, Stanford, CA 94305, USA}
\author{Michael Mulligan}
\affiliation{\small \it Department of Physics and Astronomy, University of California,
Riverside, CA 92511, USA}
\affiliation{\small \it Stanford Institute for Theoretical Physics, Stanford University, Stanford, CA 94305, USA}
\author{Gonzalo Torroba}
\affiliation{\small \it Centro At\'omico Bariloche and CONICET, R8402AGP Bariloche, ARG}
\author{Huajia Wang}
\affiliation{\small \it Department of Physics, University of Illinois, Urbana-Champaign, IL 61801, USA}


\begin{abstract}
We study supersymmetry breaking perturbations of the simplest dual
pair of
2+1-dimensional ${\cal N}$=2 supersymmetric field theories -- the free chiral multiplet and
${\cal N} = 2$ super-QED with a single flavor.  We find  
dual descriptions of a phase diagram containing four distinct massive phases.  
The equivalence of the intervening 
critical theories gives rise to several non-supersymmetric avatars of mirror symmetry: we find dualities relating scalar QED to a free fermion and Wilson-Fisher theories to both scalar and fermionic QED. 
Thus, mirror symmetry can be viewed as the multicritical parent duality from which these non-supersymmetric dualities directly descend.
\end{abstract}

\maketitle


\section{Introduction}\label{sec:intro}

Duality plays a central role in the modern understanding of quantum field theory.  In some cases, as with S-duality of maximally supersymmetric four-dimensional (4d) Yang-Mills theory, it refers to an exact symmetry exchanging strong and weak coupling limits of the same theory. 
In others, as with the dualities of ${\cal N}=1$ supersymmetric field theories in four dimensions \cite{Seibergduality} or mirror symmetry of 3d supersymmetric gauge theories \cite{Intriligator:1996ex, Aharony, Kapustin:1999ha}, it relates the low-energy physics arising from two distinct high energy theories.  Dualities have found diverse applications in high energy physics, condensed matter physics, and mathematics.

An important way to deepen our understanding of duality is to relate one duality to another.  This simplifies the logical structure of the assumptions we must make, yielding a web of dualities from a single starting point, and can also allow us to derive new dualities.

Motivated in part by the proposal \cite{Son2015} for a duality governing the physics of the half-filled Landau level, there have been several recent discussions for dualities relating some of the simplest non-supersymmetric 3d field theories \cite{Aharony2016, Karch:2016sxi,Seiberg:2016gmd,
2016arXiv160601912M,HsinSeiberg2016,awesome} (with closely related earlier work appearing in  \cite{GMPTWY2012, AharonyGurAriYacoby2012, AharonyGurAriYacobysecond2012, WangSenthilfirst2015,
MetlitskiVishwanath2016,Kachru:2015rma,Geraedtsetal2015,XuYou2015selfdual,
MrossAliceaMotrunichexplicitderivation2016, MrossAliceaMotrunichbosonicph2016}).  In this note, we show that many of these dualities
can be derived as a consequence of the most basic avatar of mirror symmetry of 3d ${\cal N}=2$ supersymmetric gauge theories.\footnote{On a related direction, it is also possible to deform three-dimensional Seiberg duality \cite{benini11,Dimofte2011, Beem2012, Intriligator:2013lca} to derive non-supersymmetric dualities in large N limits, as was done in \cite{JainMinwallaYokoyama2013, Gur-AriYacoby2015}.} We build on our work~\cite{awesome} to show that, after breaking supersymmetry, the duality between the theory of a free chiral multiplet -- theory A -- and supersymmetric quantum electrodynamics (SQED) with a single flavor -- theory B -- yields a rich phase diagram with four distinct phases.  Duality relates both the phases and the intervening critical points of the dual pairs.  Along the four walls separating distinct phases (see Figs. \ref{fig:thA} and \ref{fig:thB}), we find critical theories with dual descriptions realized in the A and B pictures.   This unifies the simplest duality of 3d supersymmetric field theory with various dualities relating fermionic and scalar QED to theories of free fermions or Wilson-Fisher bosons.  It provides a logical completion of~\cite{awesome}, where the duality between phases I and II (and the intervening critical point) was already derived.

The dualities studied here are of interest both for their intrinsic importance in understanding the structure of 3d quantum field theory, and for potential applications to problems in condensed matter physics including the study of topological order and metallic criticality.

\section{Chiral mirror symmetry}\label{sec:chiral}

We first review the essential properties of the chiral mirror symmetry duality \cite{awesome, Tong:2000ky}. 

Theory A consists of a free chiral superfield $(v, \Psi)$ which contains a complex scalar $v$ and its (two-component) Dirac fermion superpartner $\Psi$. 
The theory enjoys two global abelian symmetries, $U(1)_{J}$ and $U(1)_R$, whose actions on $(v, \Psi)$ are given below.
\be\label{eq:chiraltab1A}
\begin{tabular}{c|cc}
& $U(1)_J$ & $U(1)_R$ \\
\hline
&&\\[-8pt]
$v$ & 1 & 1  \\
&&\\[-8pt]
$\Psi$ & 1 & 0  \\
\end{tabular}
\ee
Introducing background gauge fields $\hat{A}_{J,R}$ associated to the global $U(1)_{J, R}$ symmetries, the theory A Lagrangian,
\begin{align}
\label{eq:LA-backgrounds}
\mc L^{(A)} & = |D_{\hat{A}_J  + \hat{A}_R} v|^2 - m^2_v |v|^2 \cr
& + \bar \Psi i \slashed{D}_{\hat{A}_J} \Psi - m_\Psi  \bar \Psi \Psi - {1 \over 8 \pi} \hat{A}_J d \hat{A}_J\,.
\end{align}
For abelian gauge fields $A$ and $B$, the covariant derivatives $D_{\pm A} \equiv \partial_\mu \mp i A_\mu$ with $\mu \in \{0,1,2\}$; $\slashed{D}_B \equiv \gamma^\mu (\partial_\mu - B_\mu)$ and $\bar{\Psi} \equiv \Psi^\dagger \gamma^0$ with $\gamma$-matrices\footnote{We choose the metric $\eta^{\mu \nu} = {\rm diag}(1, -1, -1)$ and $\gamma$-matrices that satisfy $(\gamma^0 \gamma^1 \gamma^2)_{\alpha \beta} = - i \delta_{\alpha \beta}$, e.g., $\gamma^0 = \sigma^3, \gamma^1 = i \sigma^1, \gamma^2 = i \sigma^2$ where $\sigma^j$ are the Pauli-$\sigma$ matrices.}
 satisfying $\{\gamma^\mu, \gamma^\nu \} = 2 \eta^{\mu \nu}$.
Chern-Simons (CS) terms are written as $A d B \equiv \epsilon^{\mu \nu \rho} A_\mu \partial_\nu B_\rho$ with $\epsilon^{012} = 1$.\footnote{We denote the effect of integrating out a single Dirac fermion of unit charge and mass $m$ by the level-$1/2$ CS term ${{\rm sgn}(m) \over 8 \pi} A d A$ in the Wilsonian effective Lagrangian. A gauge-invariant expression utilizes the eta-invariant -- see \cite{AlvarezGaume:1984nf, Wittenfermionpathintegrals2016, Seiberg:2016gmd} for further information.} Besides the relevant bosonic and fermionic mass terms, theory A admits a relevant perturbation proportional to $|v|^4$ (which will play a key role below), and a classically marginal interaction $|v|^2\bar\Psi \Psi $.

The $\hat{A}_{J}$ gauge field can be included in a background vector multiplet ${\cal{\hat{V}}}_J = (\hat{A}_{J}, \hat{\sigma}_{J}, \hat{\lambda}_{J}, \hat{D}_{J})$ with the couplings occurring in \eqref{eq:LA-backgrounds} dictated by unbroken ${\cal N}=2$ supersymmetry:
\be
\label{theoryAmasses}
m^2_v =  \hat{\sigma}_J^2 + \hat{D}_J\;\;,\;\;
m_\Psi =  \hat{\sigma}_J\,.
\ee
Chiral mirror symmetry allows us to map all components in this multiplet across the duality.
The scalar $\hat \sigma_J$ and D-term $\hat D_J$, in particular, play important roles in our derivation.

Theory B is ${\cal N}=2$ SQED with a single chiral flavor $(u, \psi)$ and $U(1)_a$ gauge group. The corresponding charge assignments are given below.
\be\label{eq:thBchargescartanchiral}
\begin{tabular}{c|ccc}
& $U(1)_J$ &  $U(1)_R$ & $U(1)_a$ \\
\hline
&&&\\[-8pt]
$u$ & 0 & 0 & -1 \\
&&&\\[-8pt]
$\psi$ & 0  & -1 & -1\\
&&& \\[-8pt]
$e^{2\pi i \gamma/g^2}$  & 1& 0 & 0 \\
&&&\\[-8pt]
$\sigma$  & 0& 0 & 0 \\
&&&\\[-8pt]
$\lambda$  & 0& -1 & 0 \\
\end{tabular}
\ee
$\sigma$ is the real scalar partner of the dynamical 3d gauge field $a_\mu$, $\lambda$ is the gaugino\footnote{$\lambda$ is the charge-conjugate of the fermion that appears in the conventionally-defined ${\cal N}=2$ vector multiplet.}, and $\gamma$ is the dual photon ($\partial^\mu \gamma = {1 \over 2 \pi} \epsilon^{\mu \nu \rho} \partial_\nu a_\rho$). 
Including the background $U(1)_{J,R}$ terms considered previously, the dual theory B Lagrangian:
\begin{align}
\mc L^{(B)}= \mc L_V + \mc L_\text{matter} + \mc L_{CS} - \mc L_{BF}
\end{align}
with
\bea
\label{lagrangiancomponents}
\mc L_V&= & \frac{1}{g^2} \left(- \frac{1}{4} f_{\mu\nu}^2 + \frac{1}{2} ( \partial \sigma)^2+ \bar \lambda i \slashed{ D}_{- \hat A_R} \lambda+ \frac{1}{2} D^2 \right), \nonumber\\
\mc L_\text{matter}& = & |D_{-a} u|^2 + \bar \psi i \slashed{D}_{-a - \hat A_R} \psi - ( \sigma^2 - D ) |u|^2 \nonumber\\
& + &\sigma \bar \psi \psi + u^* \bar{\lambda} \psi + u \bar{\psi} \lambda\,, \nonumber\\
\mc L_{CS}&= & \frac{1}{8\pi} (a da+2 D \sigma + \bar \lambda \lambda), \\
\mc L_{BF}&= & {1 \over 2 \pi} \Big(a d \hat A_J+ \hat D_J \sigma + \hat \sigma_J D \Big)+ {1 \over 4 \pi} a d \hat A_R \,.\nonumber
\eea
$f_{\mu \nu} = \partial_\mu a_\nu - \partial_\nu a_\mu$ is the field strength of the $U(1)_a$ gauge field.

Chiral mirror symmetry says that the IR ($g \rightarrow \infty$) limit of theory B has the free field description given by theory A. 

In order to understand the RG flows below, let us now discuss the map of relevant deformations. $\hat \sigma_J$ and $\hat D_J$ give rise to masses in theory A via (\ref{theoryAmasses}). These parameters are external backgrounds for the $U(1)_J$ supercurrent, and are mapped exactly to theory B according to the last line in (\ref{lagrangiancomponents}). If we further restrict to high energies, theory B is weakly coupled and $\sigma$ and $D$ may be integrated out to show that $\hat \sigma_J$ and $\hat D_J$ produce masses for $u$ and $\psi$. The other renormalizable deformations of theory A, $|v|^4$ and $|v|^2 \bar \Psi \Psi$, do not correspond to conserved currents, and hence their effect in theory B is more involved. Nevertheless, we may derive an approximate correspondence by noting that $v^* \Psi \sim \lambda$ from the quantum numbers in both theories, and $|v|^2 \sim \sigma$ from the couplings of $\hat D_J$ on both sides. These also agree with the Taub-NUT map of the underlying $\mc N=4$ theory \cite{SachdevYin, Hook:2014dfa}. Therefore, $|v|^4$ and $|v|^2 \bar \Psi \Psi$ map to masses for $\sigma$ and the gaugino in theory B.
We stress that this map is approximate and is expected to receive large quantum corrections in the IR.


\section{Dynamics and phases of theory A}\label{sec:thA}

\begin{figure}[h!]
\centering
\includegraphics[width=.9\linewidth]{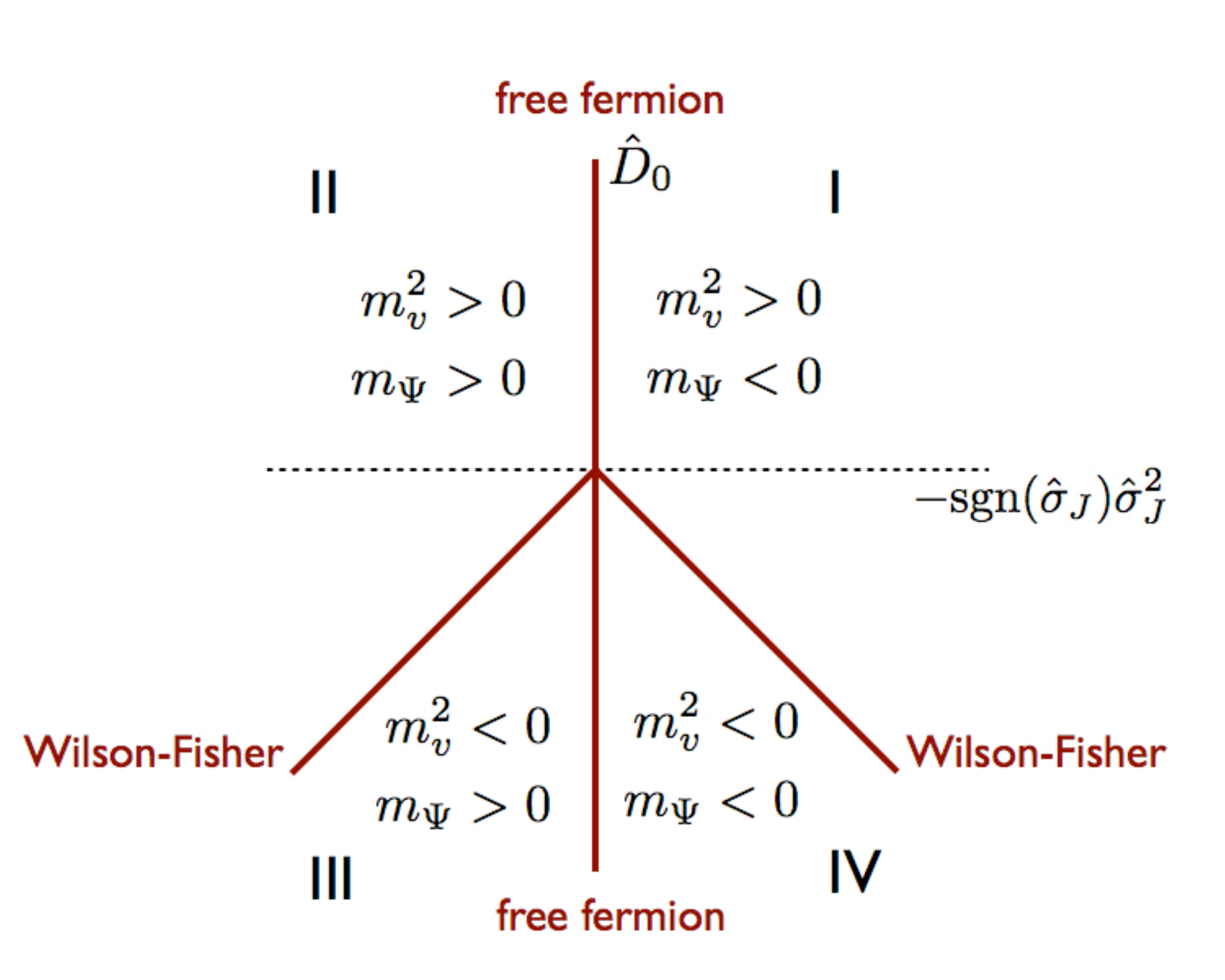}\\
\caption{Phase diagram of theory A. Massless fields occur along the red lines (second order phase transitions).}
\label{fig:thA}
\end{figure}

Theory A has four distinct phases parameterized by the signs of the effective masses $m_v^2$ and $m_\Psi$ -- see Fig. \ref{fig:thA}.
These gapped phases can be invariantly characterized by their responses to the background $\hat{A}_{J,R}$ fields.\footnote{This statement is slightly imprecise as the response can be modified by regularization-dependent contact terms \cite{ClossetDumitrescuFestucciaKomargodskiSeiberg}; the difference of the response across a phase transition is, however, physical and this is how our expressions should be understood.}
Integrating out the massive degrees of freedom according to the prescription in Fig. \ref{fig:thA}, this response is captured by the effective Lagrangians:
\begin{align}
\label{effectiveresponse}
{\cal L}^{\rm I} & = - {1 \over 4 \pi} \hat{A}_J d \hat{A}_J, \cr
{\cal L}^{\rm II} & = 0, \cr
{\cal L}^{\rm III} & = - {1 \over 2 \pi} b d (\hat{A}_J + \hat{A}_R), \cr
{\cal L}^{\rm IV} & = - {1 \over 4 \pi} \hat{A}_J d \hat{A}_J - {1 \over 2 \pi} b d (\hat{A}_J + \hat{A}_R).
\end{align}
For simplicity, we have truncated the effective Lagrangians at the leading quadratic order in the derivative expansion.
In ${\cal L}^{{\rm III}}$ and ${\cal L}^{{\rm IV}}$, we introduced the 3d gauge field $b$ \cite{MaldacenaMooreSeiberg2001} whose equation of motion imposes  the symmetry-breaking constraint,
\begin{align}
\hat{A}_J + \hat{A}_R = 0,
\end{align}
that occurs when $m^2_v < 0$.

Phases I and II, along with the intervening critical point, were studied in \cite{awesome}.
To reliably study phases III and IV, where the effective mass-squared $m_v^2 < 0$, we introduce a stabilizing interaction $|v|^4$.
Even though this interaction breaks supersymmetry, it does not qualitatively affect the analysis or conclusions for the phase structure when $m_v^2 > 0$.
It does modify the precise location at which $m_v^2 = 0$:
the classical location $\hat{\sigma}_J^2 + \hat{D}_J = 0$ -- see \eqref{theoryAmasses} -- is modified quantum mechanically.

To add the $|v|^4$ interaction in a way that can be tracked across the duality, we promote $\hat{D}_J \rightarrow D_J$ to a dynamical field and integrate it out with Gaussian weight ${\cal L}_{D_J} = {1 \over 2 h^2} (\hat{D}_J - \hat{D}_0)^2$.
(This technique was used in \cite{Gur-AriYacoby2015} in a different context.) The resulting classical scalar potential,
\be
V_\text{eff}= (\hat \sigma_J^2 + \hat D_0) |v|^2 + \frac{h^2}{2} |v|^4,
\ee
features a stable vacuum at $|v|^2= (|\hat D_0| - \hat \sigma_J^2)/h^2$ 
when $\hat D_0<- \hat \sigma_J^2$. 
The phase diagram in Fig. \ref{fig:thA} contains four lines of phase transitions: these transitions are described either by critical Wilson-Fisher theories with interaction strength determined by $h^2$ or by the mass sign-changing transition of a free Dirac fermion.

\section{Dynamics and phases of theory B}\label{sec:thB}

Direct analysis of theory B for all values of the background parameters is subtle because the theory is strongly coupled.
Duality and supersymmetry (when unbroken), however, may together be used to determine the theory B dynamics.

\begin{figure}[h!]
\centering
\includegraphics[width=.9\linewidth]{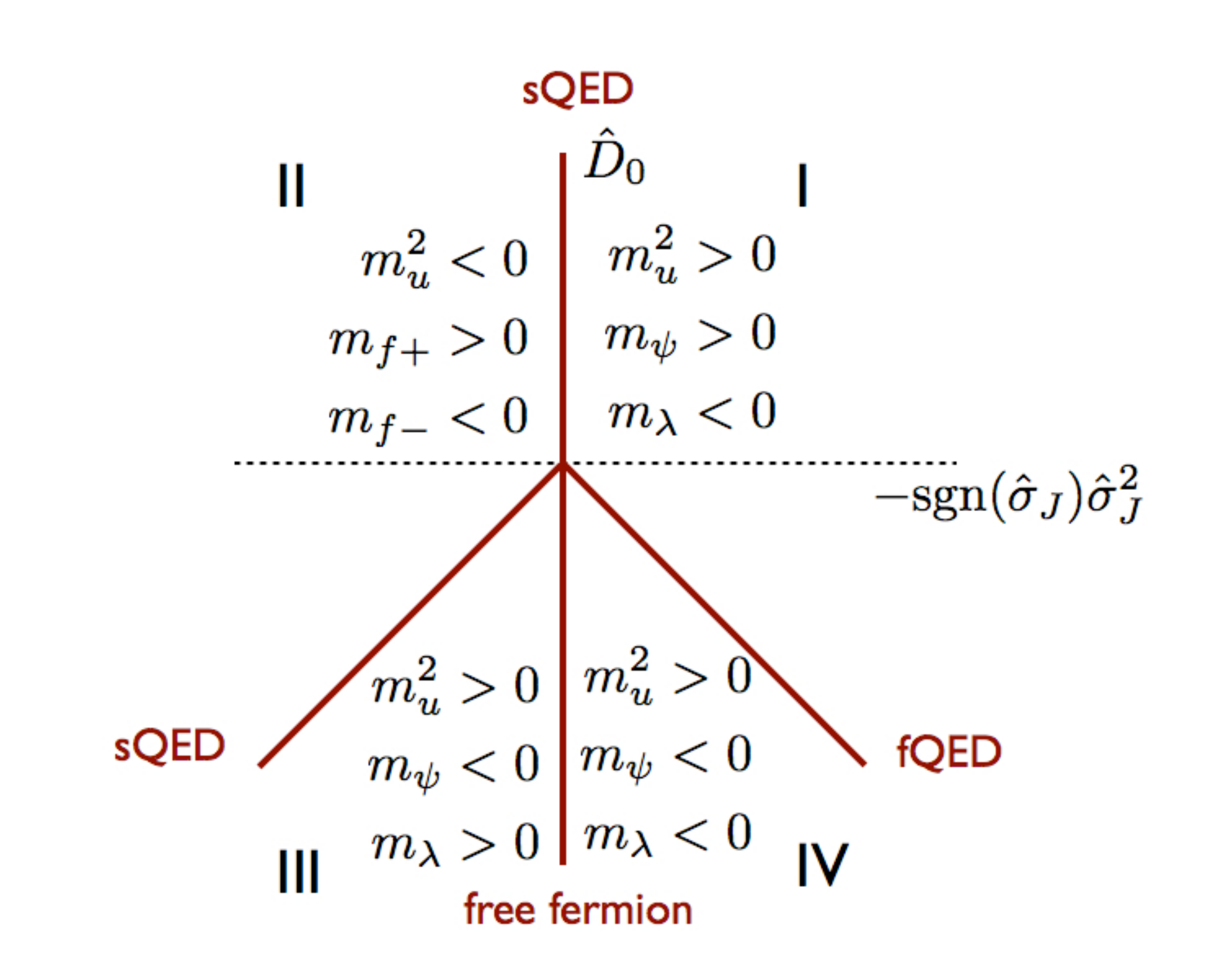}\\
\caption{Phase diagram of theory B.}
\label{fig:thB}
\end{figure}

We have found it possible to uniquely realize the theory A response summarized in \eqref{effectiveresponse} using the effective mass parameters $m_u^2$ and $m_{f_\pm}$ for the charged degrees of freedom of theory B with the prescription given in Fig. \ref{fig:thB}.
We have introduced mass parameters for Dirac fermions $f_{\pm}$ that diagonalize the effective fermion mass matrix in
\begin{align}
\label{massmatrix}
{\cal L}^{(B)}_{\rm mass} = - m_\psi \bar{\psi} \psi - m_\lambda \bar{\lambda} \lambda + \delta m^\ast\, \bar{\lambda} \psi + \delta m \,\bar{\psi} \lambda.
\end{align}
The third and fourth terms -- see the second line of ${\cal L}_{\rm matter}$ -- are parameterized by the fermion mass mixing $\delta m$ which is only non-zero in phases when $\langle |u^2| \rangle \neq 0$, i.e., $m_u^2 < 0$.
The mass matrix in \eqref{massmatrix} has the two eigenvalues,
\begin{align}
\label{masseigenvalues}
m_{f_{\pm}} = {1 \over 2} \Big(m_\psi + m_\lambda \pm \sqrt{(m_\psi - m_\lambda)^2 + 4 |\delta m|^2} \Big).
\end{align}
$f_\pm$ are composed of linear combinations of $\psi$ and $\lambda$ with relative weights that vary as the background parameters are tuned: Fig. \ref{fig:thB} implies that upon transitioning from phase II to phase I, $f_+$ becomes $\psi$ and $f_-$ becomes $\lambda$ and vice versa in the transition from phase II to phase III.

The matching of the theory B responses to those of theory A in \eqref{effectiveresponse} follows upon integrating out the massive degrees of freedom,
\begin{align}
\label{theoryBeffresponse}
{\cal L}_{\rm eff}^{(B)} & = {1 \over 8 \pi} {\rm sgn}(m_{f_+}) (a + \hat{A}_R) d (a + \hat{A}_R) + {1 \over 8 \pi} a d a \cr 
& +  {1 \over 8 \pi} {\rm sgn}(m_{f_-}) \hat{A}_R d \hat{A}_R - {1 \over 4 \pi} a d \hat{A}_R - {1 \over 2 \pi} \hat{A}_J d a.\cr
\end{align}
This formula should be read with the replacement $f_+ = \psi$ and $f_- = \lambda$ in phases where $m^2_u>0$.
In phases I, III, and IV where $m_u^2 > 0$, the response in \eqref{theoryBeffresponse} is seen to directly match that in \eqref{effectiveresponse} for the values of the effective mass parameters given in Fig. \ref{fig:thB}.
In phase I, the CS $a da$ coefficient is nonzero, and integrating out $a$ matches the result from theory A. On the other hand, in phases III and IV, this coefficient vanishes and $a$ is identified with the gauge field $b$ introduced in (\ref{effectiveresponse}); this reproduces the correct symmetry breaking pattern and CS response.
In phase II, the condensation $\langle |u|^2 \rangle \neq 0$ essentially fixes $a =0$ and we find agreement with ${\cal L}^{\rm II}$ in \eqref{effectiveresponse} after using the effective mass values in Fig. \ref{fig:thB}.
The response in phase II requires $m_{f_+} m_{f_-} < 0$ or
\begin{align}
m_\psi m_\lambda < |\delta m^2|\,.
\end{align}
Happily, this requirement is consistent with the values of the effective masses in the adjacent phases I and III.

To provide additional justification for this picture, we analyze theory B near the line $\hat{D}_0 = 0$ where supersymmetry is preserved.
Because supersymmetry precludes phase transitions as a function of the coupling $g$, we may transfer the qualitative information gleaned at weak coupling to the strong coupling regime of interest.
Note that we are ignoring the effects of the dynamical $D_J$ field here; it provides a stabilizing potential for $\sigma$, but can otherwise be ignored near the origin of Fig. \ref{fig:thB}.

We first verify the effective mass parameter assignments in phases I and II along $\hat{D}_0 = 0$.
In \cite{awesome}, we demonstrated that the supersymmetry-preserving vacuum for the scalar fields lies at
\begin{align}
\label{susyvacua}
\langle |u|^2 \rangle & = 0, & \langle \sigma \rangle & = \hat{\sigma}_J, & {\rm for}\ & \hat{\sigma}_J < 0,\cr
\langle |u|^2 \rangle & = {\hat{\sigma}_J \over 2 \pi}, & \langle \sigma \rangle & = 0, & {\rm for}\ & \hat{\sigma}_J > 0. 
\end{align}
From this analysis we abstract the following.
For $\hat{\sigma}_J < 0$, we determine the effective masses $m^2_u > 0$, $m_\psi = - \langle \sigma \rangle > 0$, and $m_\lambda < 0$ from ${\cal L}_{\rm CS}$ in \eqref{lagrangiancomponents}.
For $\hat{\sigma}_J > 0$, we find that $m_u^2 < 0$, $m_\psi = 0$, and $m_\lambda < 0$.
Using the mass matrix eigenvalues in \eqref{masseigenvalues}, we find agreement with the inferred values in Fig. \ref{fig:thB}.

Perturbation theory in $\hat{D}_0 < 0$ is likewise consistent with the assignment of effective masses.
To leading order in $\hat{D}_0$, the vacua in \eqref{susyvacua} are shifted as follows:
\begin{align}
\label{perturbationresults}
\delta \langle u \rangle & = 0, & \delta \langle \sigma \rangle & \sim g^2 |\hat{D}_0| \hat{\sigma}_J^{-2}, & {\rm for}\ & \hat{\sigma}_J < 0,\cr
\delta \langle u \rangle & \sim - |\hat{D}_0| \hat{\sigma}_J^{-3/2}, & \delta \langle \sigma \rangle & \sim |\hat{D}_0| \hat{\sigma}_J^{-1}, & {\rm for}\ & \hat{\sigma}_J > 0,
\end{align}
where $\sim$ indicates equality up to multiplication by a positive constant.
For $\hat{\sigma}_J < 0$, we see that $m_\psi$ is decreased to leading order, consistent with the putative fermion mass sign-changing transition at $\hat{D}_0 = - {\rm sgn}(\hat{\sigma}_J) \hat{\sigma}_J^2$.
For $\hat{\sigma}_J > 0$, the perturbative decrease of $\langle u \rangle$ is consistent with symmetry restoration across the $\hat{D}_0 = - {\rm sgn}(\hat{\sigma}_J) \hat{\sigma}_J^2$ line.
Likewise, the perturbative decrease of $m_\psi = - \langle \sigma \rangle$ matches the expected behavior in phase III. 
Integrating out $(u,\psi)$ near the origin in phases III and IV generates a 1-loop correction 
with the result, $m_\lambda\sim |m_\psi|-|m_u|$. This agrees with the sign assignments for $m_\lambda$ near the II/III and IV/I phase boundaries, where $u$ and $\psi$ are becoming massless. 
 
\section{Dualities and implications}\label{sec:dualities}

Examination of Figs. \ref{fig:thA} and \ref{fig:thB} shows that a single field becomes light at a given phase transition.
Mirror symmetry implies the resulting critical theories are dual.
We thus arrive at the following dualities (indicated by $\leftrightarrow$): 
\begin{widetext}
\begin{align}
\label{fermionsqed}
& \bar{\Psi} i \slashed{D}_{\hat{A}_J} \Psi - {1 \over 8 \pi} \hat{A}_J d \hat{A}_J \leftrightarrow |D_{-a} u|^2 - |u|^4 + {1 \over 4 \pi} a d a - {1 \over 2 \pi} \hat{A}_J d a, \\
\label{wfsqed}
& |D_{\hat{A}_J + \hat{A}_R}v|^2 - |v|^4 \leftrightarrow -\frac{1}{4g^2}f_{\mu\nu}^2+ |D_{-a} u|^2 - |u|^4 - {1 \over 2 \pi} a d (\hat{A}_J + \hat{A}_R), \\
\label{trivialfermionduality}
& \bar{\Psi} i \slashed{D}_{\hat{A}_J} \Psi - {1 \over 8 \pi} \hat{A}_J d \hat{A}_J - {1 \over 2 \pi} b d (\hat{A}_J + \hat{A}_R) \leftrightarrow \bar{\lambda} i \slashed{D}_{- \hat{A}_R} \lambda - {1 \over 8 \pi} \hat{A}_J d \hat{A}_J - {1 \over 2 \pi} a d (\hat{A}_J + \hat{A}_R), \\
\label{wffqed}
& |D_{\hat{A}_J + \hat{A}_R}v|^2 - |v|^4 - {1 \over 4 \pi} \hat{A}_J d \hat{A}_J \leftrightarrow \bar{\psi} i \slashed{D}_{- a - \hat{A}_R} \psi + {1 \over 8 \pi} a d a - {1 \over 2 \pi}a d (\hat{A}_J + {1 \over 2} \hat{A}_R) - {1 \over 8 \pi} \hat{A}_R d \hat{A}_R.
\end{align}
\end{widetext}
\eqref{fermionsqed} and \eqref{wffqed} are two examples of the bosonization dualities discussed in \cite{Aharony2016, Seiberg:2016gmd}.
The supersymmetric chiral mirror duality can be viewed as the multicritical parent duality from which these non-supersymmetric dualities descend.

\eqref{fermionsqed} provides dual descriptions for the integer quantum Hall plateau transition between phases I and II where the level of the $\hat{A}_J$ CS term changes by unity.
\eqref{wfsqed} is Peskin-Dasgupta-Halperin duality \cite{Peskin:1977kp, DasguptaHalperin1981} and may be used to describe a superfluid transition where the diagonal component of the $U(1)_J \times U(1)_R$ symmetry is broken.
\eqref{trivialfermionduality} is a ``trivial" duality relating a free fermion of theory A to a free fermion of theory B.
\eqref{wffqed} again describes a $U(1)_J \times U(1)_R$ symmetry-breaking transition; the distinction from \eqref{wfsqed} lies in the presence of a level-1 CS term for $\hat{A}_J$ in the adjacent massive phases.

In \cite{Karch:2016sxi, Seiberg:2016gmd}, it was shown how the action of the modular group on dual conformal field theories (CFTs) can generate additional dualities.
Here, we discuss how this action relates the above dualities to one another.
Denoting the Lagrangian of a general CFT by ${\cal L}(\Phi, \hat{A})$, where $\Phi$ collectively represents the fields of the CFT and $\hat{A}$ is a background field for the global $U(1)$ symmetry, the modular group acts as follows \cite{WittenSL2Z2003, LeighPetkouSL2Z2003}:
\begin{align}
\label{CFTaction}
& {\cal T}: {\cal L}(\Phi, \hat{A}) \mapsto {\cal L}(\Phi, \hat{A}) + {1 \over 4 \pi} \hat{A} d \hat{A}, \cr
& {\cal S}: {\cal L}(\Phi, \hat{A}) \mapsto {\cal L}(\Phi, a) - {1 \over 2 \pi} \hat{B} d a. 
\end{align}
The action ${\cal S}$ deserves further explanation: this transformation makes the background field dynamical $\hat{A} \rightarrow a$ and adds to the Lagrangian a new background field via the BF coupling $- {1 \over 2 \pi} \hat{B} d a$.
The rules in \eqref{CFTaction} induce the action ${\cal T}: \t \sigma \mapsto \begin{pmatrix}1 & 1 \cr 0 & 1 \end{pmatrix} \t \sigma$ and ${\cal S}: \t \sigma \mapsto \begin{pmatrix}0 & 1 \cr -1 & 0 \end{pmatrix} \t \sigma$ on the complexified conductivity $\t \sigma = \sigma_{xy} + i \sigma_{xx} \in \mathbb{H}$ (extracted from the two-point functions of the $U(1)$ current).

Using \eqref{CFTaction}, we see that \eqref{wffqed} is the ${\cal S}$-transform of \eqref{fermionsqed}.
Examining the sides of the initial duality that a given theory occurs, we may say that ${\cal S}$ exchanges theory A and B in a loose sense.
Performing a ${\cal T}$ transformation on \eqref{wffqed}, we obtain a fermionic description for the $U(1)_J \times U(1)_R$ symmetry-breaking transition complementary to \eqref{wfsqed}.
(A closely related duality to \eqref{wffqed} was previously argued for in \cite{Mulligan2016}.)
Using the marvelous property of transitivity, we can relate the right-hand side of \eqref{wfsqed} to the ${\cal T}$-transform of the right-hand side of \eqref{wffqed}.




\acknowledgements
This research was supported in part by 
grant NSF PHY-1316699 (S.K.), the University of California (M.M.), Conicet PIP-11220110100752 (G.T.), and DARPA YFA contract D15AP00108 (H.W.).
M.M. is grateful for the generous hospitality of the Aspen Center for Physics NSF PHY-1066293 and the Kavli Institute for Theoretical Physics NSF PHY-11-25915.

\bibliography{QHE.bib}{}
\bibliographystyle{utphys}

\end{document}